\documentclass{article}
\usepackage[margin=1in]{geometry}

\usepackage[authoryear,round,sort]{natbib}
\usepackage{microtype}
\usepackage{authblk}
\usepackage{amsmath}
\usepackage{amssymb}
\usepackage{mathtools}
\usepackage{graphicx}
\usepackage{bm}
\usepackage[dvipsnames]{xcolor}
\usepackage[colorlinks=true,allcolors=NavyBlue]{hyperref}
\usepackage{newpxtext,newpxmath}

\defcitealias{zv2021activation}{ZP}

\usepackage{lineno}

\begin{document}

\title{On neural network kernels and the storage capacity problem}

\author[1,2]{Jacob A. Zavatone-Veth\thanks{\href{mailto:jzavatoneveth@g.harvard.edu}{jzavatoneveth@g.harvard.edu}}}
\author[2,3]{Cengiz Pehlevan\thanks{\href{mailto:cpehlevan@seas.harvard.edu}{cpehlevan@seas.harvard.edu}}}

\affil[1]{Department of Physics, Harvard University, Cambridge, MA 02138}
\affil[2]{Center for Brain Science,  Harvard University, Cambridge, MA 02138}
\affil[3]{John A. Paulson School of Engineering and Applied Sciences,  Harvard University, Cambridge, MA 02138}

\date{\today}

\maketitle

\begin{abstract}
    In this short note, we reify the connection between work on the storage capacity problem in wide two-layer treelike neural networks and the rapidly-growing body of literature on kernel limits of wide neural networks. Concretely, we observe that the ``effective order parameter'' studied in the statistical mechanics literature is exactly equivalent to the infinite-width Neural Network Gaussian Process Kernel. This correspondence connects the expressivity and trainability of wide two-layer neural networks. 
\end{abstract}

The study of two-layer neural networks in the limit of large hidden layer width has a long history in the statistical physics of learning \citep{barkai1992broken,engel1992storage}. This work focuses on the Gardner storage capacity problem, which measures the computational power of a neural network by the largest random binary classification dataset it can ``memorize'' \citep{gardner1988space,gardner1988optimal}. In this problem, one performs Bayesian inference of network weights using a likelihood that is flat on the set of weights that yield zero classification error and vanishes uniformly otherwise. Memorization is possible if the support of the resulting Bayes posterior has nonzero volume, i.e., if there exists a non-negligible set of weights for which all examples are correctly classified. In the thermodynamic limit where the input dimensionality $N$ and dataset size $P$ tend to infinity with fixed ratio $\alpha \equiv P/N$, there is a sharp transition between memorization with probability one and memorization with probability zero at some critical ratio $\alpha_{c}$, referred to as the storage capacity. 

These statistical mechanics calculations rely on studying functions of the overlap $q$ between the hidden unit weight vectors of identical copies, known as replicas, of the network, which is termed the ``order parameter.'' Classic work by \citet{barkai1992broken} and \citet{engel1992storage} for treelike networks with sign function hidden layer activations revealed a remarkable simplification in the limit where the number of hidden units $K$ was taken to be large: the equations reduced to those of a perceptron \citep{gardner1988space,gardner1988optimal}, with the order parameter replaced by an ``effective'' order parameter $q_{\textrm{eff}}(q)$. This result was recently extended to rectified linear unit activation functions by \citet{baldassi2019properties} and to general activation functions in our own work \citep[][hereafter \citetalias{zv2021activation}]{zv2021activation}.

Beginning with pioneering work by \citet{neal1996priors} and \citet{williams1997computing}, a parallel line of research in the machine learning community has characterized the infinite-width limits of Bayesian neural networks with Gaussian priors over their weights. This work also revealed a remarkable simplification: for a two-layer network with hidden layer activation function $f$, infinite-width inference is equivalent to shallow Gaussian process (GP) inference with kernel\footnote{For clarity of exposition, we make the simplifying assumptions that the network has no bias terms and that the prior weight variance is unity.}
\begin{linenomath*}
\begin{equation} \label{eqn:nngpkernel}
    K_{f}(\mathbf{x},\mathbf{y}) = \mathbb{E}\left[ f(\mathbf{w} \cdot \mathbf{x}) f(\mathbf{w} \cdot \mathbf{y}) \,:\, \mathbf{w} \sim \mathcal{N}(\mathbf{0},\mathbf{I}) \right]
\end{equation}
\end{linenomath*}
given by the deterministic infinite-width limit of the Gram matrix of hidden layer activations. Here, $\mathbf{x},\mathbf{y} \in \mathbb{R}^{N}$ are interpreted as two inputs to the network, and $\mathbf{w} \sim \mathcal{N}(\mathbf{0},\mathbf{I})$ is a random weight vector. In recent years, this neural network-Gaussian process (NNGP) correspondence has proven to be an extremely fruitful tool in the theory of deep learning thanks to the observation that it can be extended to networks with more than a single hidden layer \citep{lee2018deep,matthews2018gaussian}. Importantly, this correspondence does not apply only at the level of the prior, but also at the level of the Bayes posterior. For sensible likelihoods that model the targets as being conditionally independent of the network weights given the network outputs, the function-space posterior distribution tends to that induced by the limiting Gaussian process prior \citep{hron2020exact}. Moreover, the connection between infinitely-wide neural networks and kernel machines has been extended to study gradient-based training \citep{jacot2018ntk}. 

However, a precise connection between these two lines of research on Bayesian inference in infinitely-wide two-layer neural networks has yet to be drawn. In this brief note, we re-interpret results on the storage capacity problem in terms of the behavior of the NNGP kernel. This observation allows us to clarify the conceptual and technical connections between these parallel lines of work on wide and deep neural networks. 

We start by observing that, for unit-norm inputs $\mathbf{x},\mathbf{y} \in \mathbb{S}^{N-1}$, the infinite-width NNGP kernel \eqref{eqn:nngpkernel} can be expressed as a function of the overlap $q \equiv \mathbf{x} \cdot \mathbf{y} \in [-1,+1]$ \citep{neal1996priors,williams1997computing,cho2009kernel,lee2018deep,matthews2018gaussian}: 
\begin{linenomath*}
\begin{equation}
    K_{f}(q) = \begin{dcases} \mathbb{E}\left[ f(x)^2 \,:\, x \sim \mathcal{N}(0,1) \right], & \textrm{if } q = 1 \\
    \mathbb{E}\left[ f(x) f(y) \,:\, \begin{pmatrix} x \\ y \end{pmatrix} \sim \mathcal{N}\left(\begin{pmatrix} 0 \\ 0 \end{pmatrix}, \begin{pmatrix} 1 & q \\ q & 1 \end{pmatrix} \right) \right], & \textrm{if } -1 < q < 1 
    \\
    \mathbb{E}\left[ f(x) f(-x) \,:\, x \sim \mathcal{N}(0,1) \right], & \textrm{if } q = -1 . 
    \end{dcases}
\end{equation}
\end{linenomath*}
For functions $f$ that are square-integrable with respect to Gaussian measure, $K_{f}(q)$ is a continuous function on the closed interval $[-1,1]$, with a power series that converges throughout that interval \citep{zv2021activation,daniely2016toward,bogachev1998gaussian}. 

The NNGP kernel coincides exactly with the effective order parameter studied in the statistical mechanics literature:\footnote{Possibly up to irrelevant constant offsets; in \citetalias{zv2021activation} we defined the effective order parameter as $q_{\textrm{eff}}(q) = K_{f}(q) - [\mathbb{E}_{x \sim \mathcal{N}(0,1)}f(x)]^2$.}
\begin{linenomath*}
\begin{equation}
    q_{\textrm{eff}}(q) = K_{f}(q).
\end{equation}
\end{linenomath*}
In particular, the result of \citetalias{zv2021activation} shows that the storage capacity remains finite in the infinite-width limit if the left derivative 
\begin{linenomath*}
\begin{equation}
    \partial_{-} K_{f}(q=1) \equiv \lim_{q \uparrow 1} \frac{K_{f}(1) - K_{f}(q)}{1-q}
\end{equation}
\end{linenomath*}
of the kernel at $q=1$ is finite, and diverges otherwise. For the special cases of $f(x) = \operatorname{sign}(x)$, which yields infinite capacity, and $f(x) = \operatorname{ReLU}(x) = \max\{0,x\}$, which yields finite capacity, this was noted in previous work by \citet{barkai1992broken} and by \citet{baldassi2019properties}, respectively, based on direct computations of $K_{\operatorname{sign}}(q)$ and $K_{\operatorname{ReLU}}(q)$.\footnote{In the machine learning literature, kernels of this family were studied systematically by \citet{cho2009kernel}. } Thus, the relationship of the limiting behavior of the NNGP kernel for sign activation functions to the expressivity of infinitely-wide networks was implicitly studied thirty years ago.

Therefore, the storage capacity of a treelike committee machine is related to the behavior of the NNGP kernel for nearly colinear arguments. In \citetalias{zv2021activation}, we gave a general argument based on Fourier-Hermite expansions that $\partial_{-} K_{f}(q=1)$ is finite if and only if the activation function $f$ is in the Sobolev space of functions that are square-integrable with respect to Gaussian measure and have weak derivatives that are also square-integrable with respect to Gaussian measure \citep{bogachev1998gaussian}. Roughly speaking, a kernel with a cusp at $q = 1$ will yield divergent capacity. Intuitively, $\partial_{-} K_{f}(q=1)$ measures the ability of the kernel to discriminate between nearly colinear inputs, which has a natural relationship to expressivity \citep{paccolat2021isotropic,poole2016exponential}. 

The relationship between $\partial_{-} K_{f}(q=1)$ and the expressivity of deep networks was noted in a different context by \citet{poole2016exponential}. Those authors studied the expressivity of infinitely wide and deep fully-connected networks with random weights in terms of the fixed points of $K_{f}(q)$ under iteration. For networks without bias terms, their results have a close relationship to those of \citetalias{zv2021activation}, which we will illustrate in a simple setting. We assume that all inputs lie on the sphere, and normalize the activation function such that $K_{f}(q=1) = 1$. Then, the result of \citet{poole2016exponential} shows that networks with $\partial_{-} K_{f}(q=1) > 1$ display chaotic behavior in the sense that the fixed points of $K_{f}(q)$ are unstable, while those with $\partial_{-} K_{f}(q=1) < 1$ display ordered behavior. Those authors proposed that the sharp enhancement of differences between inputs by networks in the chaotic regime is a signature of expressivity. We remark that capacity calculation is conceptually distinct from the settings of GP inference and \citet{poole2016exponential} in that the effective order parameter measures the similarity between the weight vectors of two different replicas for a random input example, while the kernel measures the similarity of two input examples for a random weight vector.

The connection between the storage capacity problem and the NNGP kernel also helps to clarify the relationship of \citetalias{zv2021activation} to \citet{panigrahi2020effect}'s study of gradient descent training in two-layer networks with fixed readout weights. In their setting, the speed of training is governed by the minimum eigenvalue of the ``gradient Gram matrix'' evaluated on the training examples, schematically given as $G_{f}(\mathbf{x},\mathbf{y}) = \mathbf{x} \cdot \mathbf{y} K_{f'}(\mathbf{x},\mathbf{y})$.\footnote{Here, the activation function $f$ is always assumed to be at least Lipschitz continuous, and is assumed to have weak derivative $f'$ that is square-integrable with respect to Gaussian measure such that $G_{f}$ is well-defined.} Working under the assumption that all inputs lie on the sphere, they showed that activation functions with a discontinuity in their derivatives yield rapid gradient descent training under a weaker bound on the maximum overlap $q$ between any two training examples than is required for smooth activation functions. 

As we have $G_{f}(q) = q K_{f'}(q)$ for inputs on the sphere, we can apply the above intuition for the relationship between the left derivative $\partial_{-} K_{f}(q=1)$ and discriminability. By the abovementioned results, functions $f$ with discontinuous first derivatives will result in divergent left derivatives $\partial_{-} G_{f}(q=1)$ and thus sharp discrimination capabilities. This sharp discrimation capability will yield better separation between diagonal and off-diagonal elements of $G_{f}(q)$ for $q$ near one. By the Gershgorin circle theorem, this should in turn yield better lower bounds on the minimum eigenvalue of the $G_{f}$ as a function of $q$ than would hold for $G_{f}(q)$ with finite $\partial_{-} G_{f}(q=1)$, possibly decreasing training time \citep{horn2012matrix}. Therefore, the link between trainability and the second weak derivative of the activation function noted by \citet{panigrahi2020effect} and the link between storage capacity and the first weak derivative of the activation function noted in \citetalias{zv2021activation} bear a close conceptual relation: both depend on the discrimination capabilities of the appropriate kernel for nearly-colinear inputs. 


To conclude, the storage capacity of a wide two-layer treelike neural network is determined by the behavior of the corresponding NNGP kernel for nearly-colinear inputs on the sphere. This connection yields an intuitive explanation for the varying behavior of the storage capacity in terms of input discriminability \citep{barkai1992broken,engel1992storage,baldassi2019properties,zv2021activation,gardner1988space,gardner1988optimal}, as well as a more precise description of the relationship of this line of research to work on kernel limits of neural networks \citep{neal1996priors,williams1997computing,matthews2018gaussian,lee2018deep,jacot2018ntk,panigrahi2020effect,cho2009kernel,poole2016exponential,hron2020exact}. 

Yet, further research will be required to fully understand the connections between studies of the storage capacity problem in statistical physics and results on the NNGP limit. Though both lines of research focus on Bayesian inference of network weights, they consider different settings. The statistical physics literature largely focuses on simple two-layer models in a limit where the input dimension tends to infinity with the hidden layer width \citep{barkai1992broken,engel1992storage,baldassi2019properties,zv2021activation}, while most studies of the NNGP limit consider deeper fully-connected networks with finite-dimensional inputs \citep{neal1996priors,williams1997computing,matthews2018gaussian,lee2018deep,hron2020exact}. Given the growing interest in applying tools from statistical physics to study inference in wide neural networks,\footnote{See our discussion in \citet{zv2021asymptotics} and references therein.} we hope that the connections noted in this work will spark more detailed investigation of possible commonalities between these seemingly disparate settings.

\section*{Acknowledgements}

This work was supported by the Harvard Data Science Initiative Competitive Research Fund, the Harvard Dean’s Competitive Fund for Promising Scholarship, and a Google Faculty Research Award. 

\bibliography{refs.bib}

\end{document}